\documentclass[10pt,prl,twocolumn,tightenlines]{revtex4}
\usepackage{graphics}
\begin{document}
%
\title{N\'eel transition, spin fluctuations, and pseudogap 
in underdoped cuprates by 
a Lorentz invariant four-fermion model in 2+1 dimensions}
\author{Manuel Reenders\footnote{Email: m.reenders@phys.rug.nl}}
\affiliation{
Department of Polymer Chemistry and Materials Science Center,\\
University of Groningen, Nijenborgh 4,
9747 AG Groningen, The Netherlands}
\date{\today}
\begin{abstract}
We show that the N\'eel transition and spin fluctuations near 
the N\'eel transition in planar cuprates can be described by an 
SU(2) invariant relativistic four-fermion
model in 2+1 dimensions.  Features of the
pseudogap phenomenon are naturally described by the appearance of an
anomalous dimension for the spinon propagator.
\end{abstract}
\maketitle

%
%
%
The two dimensional one-band repulsive Hubbard model \cite{an87zhri88}
is considered as one of the best candidates for describing
microscopically the planar cuprate high-Tc superconductors.  The
strong coupling limit of the Hubbard model is equivalent to the t-J
model \cite{bazoan87ko88}.  In the t-J model at half-filling, on each
site of a square lattice, on average a single electron interacts
antiferromagnetically with its nearest neighbors and the system is an
antiferromagnetic (AF) insulator (a Mott insulator) described by
N\'eel ordering.  By introducing doped holes, thus charge carriers, on
this lattice, the AF interaction is frustrated and a transition from
the N\'eel ordered to the disordered, so-called spin-gapped phase or
normal state occurs.  For the Hubbard-Heisenberg model at and near
half-filling Affleck and Marston showed, using a leading-order $1/N$
expansion, that the ground state is the $\pi$-flux phase for
appropriate values of the hopping amplitude $t$, doping $\delta$, and
AF interaction $J$ \cite{afma88maaf89}.  The number $N$ is the
generalization of the physical up-down spins, $N=2$, to $N$ types.  In
the $\pi$-flux phase the spinon spectrum has the dispersion
\begin{eqnarray}
E_k\simeq 2|\chi|a\sqrt{\cos^2 a k_1+\cos^2 a k_2}, \nonumber
\end{eqnarray}
where $|\chi|$ is the absolute value of the $\pi$-flux phase order
parameter.  This spectrum is gapless at the two Fermi vectors $\vec
fp=(\pi/2a,\pm\pi/2a)$ in the reduced Brillioun zone of the even
and odd lattices with lattice spacing $a$.  The
linearization around these Fermi points gives a continuum 
(2+1)-dimensional massless Dirac theory describing $N$ flavors of four
component Dirac spinors having a global U($2N$) symmetry
\cite{afma88maaf89,sewij89}.  At half-filling, the Dirac spectrum is
isotropic and the flux-phase order parameter is equal to the so-called
d-wave pairing order parameter $|\Delta|=|\chi|$
\cite{afzohsan88dafrmo88}.

Recently Kim and Lee addressed the question, how the spin-gapped phase
is connected to the N\'eel ordered phase at zero doping \cite{kile99}.
In their work the mean-field $\pi$-flux phase of Affleck and Marston
is taken as the reference state for describing the spin fluctuations
around the AF Fermi points.  By introducing gauge field fluctuations,
enhancing AF correlations around the $\pi$-flux phase solution, Kim
and Lee propose, along the lines of Ref.~\cite{ma90a} that N\'eel
ordering is described by dynamical symmetry breaking (DSB) and
mass generation in QED$_3$.  N\'eel ordering corresponds to the
dynamically broken phase, which is characterized by a ``mass gap'' for
the spinon spectrum and Nambu-Goldstone bosons as bound states of
spinons and antispinons. These Nambu-Goldstone bosons are the massless
AF spin waves. The disordered spin-gapped phase is equivalent to the
subcritical symmetric phase and is characterized by the existence of
massless spinons and unstable bound states as broad resonances.  These
resonances supposedly correspond to the spin excitations observed in
the normal state and superconducting state of underdoped and optimally doped
cuprates \cite{fongetal}.

Although the physical picture sketched by Kim and Lee is plausible, it
was pointed out that QED$_3$ is not an appropriate model for
describing unstable bound states \cite{guhare01}.  The main problem
being that DSB in QED$_3$ is not a phase transition of the second-order 
type, but a so-called conformal phase transition, which does not
allow light unstable bound states in the symmetric phase
\cite{miya97}.  As an alternative for the gauge interactions, we
propose that relevant, Lorentz-invariant four-fermion or four-Fermi
(4F) interactions with an ultraviolet stable fixed point for the
four-fermion coupling drive the AF ordering.  The AF lattice
Heisenberg interaction $H=J\sum_{<x,y>}\vec S_x\cdot \vec S_y$,
expanded around the two Fermi points, gives rise to SU(2) invariant
attractive 4F terms in the action of the model.  At sufficient strong
AF coupling DSB occurs, giving rise to the N\'eel state.  Despite the
fact that the real temperature is not necessarily zero, the time
dependent quantum fluctuations and ordering are given by a zero
temperature 4F model.  The mean-field $\pi$-flux phase local order
parameter $\chi$ describes the thermodynamic equilibrium state and is
therefore time independent.  In addition, there is no need for a
chemical potential in the proposed model, since the (nearly)
half-filling constraint has already been taken into account via the
mean-field equilibrium real bosonic Lagrange field
\cite{afma88maaf89}.  The present idea is partly inspired by
Ref.~\cite{sewij89}, where it was suggested, by analyzing various
lattice 4F operators, that the only relevant operators are those which
are Lorentz invariant in the continuum.

We adopt the spin liquid ansatz of Refs.~\cite{afma88maaf89,kile99};
the spin liquid is described by the mean-field large-$N$ $\pi$-flux
phase for low doping. The flux-phase order parameter $|\chi|$ depends
on temperature and doping.  The AF Heisenberg interaction is
reinstated for this spin liquid.  The fluctuations of the holes are
ignored, and their effect is only included via their mean-field effect
on reducing the AF exchange $J$ to $J_e=J(1-\delta)^2$
\cite{lesa99an01}.  Therefore, on a lattice with spacing
$a^\prime=a/\sqrt{2}$, we consider the action
\begin{eqnarray}
S=\int dt\left[\sum_{<x,y>}c^\dagger_{\alpha}(x,t) (i\partial_t
-\chi_{yx})c_{\alpha}(y,t)-H_I\right],\label{lag1}
\end{eqnarray}
with
\begin{eqnarray}
H_I=\sum_{<x,y>} J_e \vec S_x\cdot \vec S_y,\label{Hheis}
\end{eqnarray}
and where $<x,y>$ denotes nearest neighbors on an isotropic cubic
lattice.  The index $\alpha=\uparrow,\downarrow$ labels the spin
components and the spin operator is
\begin{eqnarray}
\vec S_x=c^\dagger_\alpha (x,t)\vec \sigma_{\alpha\beta} c_\beta(x,t)/2, 
\nonumber
\end{eqnarray}
where $\sigma$ are the Pauli matrices and $c$, $c^\dagger$ are the
spinon, antispinon operators.  A particular representation for this
Hermitian $\pi$-flux-phase hopping parameter $\chi_{yx}$ is
\cite{afma88maaf89}
\begin{eqnarray}
\chi_{x\pm a_1,x}=i|\chi|,\quad
\chi_{x\pm a_2,x}=1|\chi|, \nonumber
\end{eqnarray}
with the nearest neighbor vectors $\vec a_1=(a^\prime,0)$ and $\vec
a_2=(0,a^\prime)$.  The low-energy behavior of the kinetic term in
Eq.~(\ref{lag1}) is known to be equivalent to a two-flavor massless
Dirac theory with the action
\begin{eqnarray}
S_k=\int dt\int\limits_{k\leq \Lambda}d^2k\,
\bar\psi_\alpha[i\partial_t\gamma^0+c (k_1\gamma^1+k_2\gamma^2)]
\psi_\alpha,\label{Lkdef}
\end{eqnarray}
where $c=2|\chi|a$ is the ``speed of light'' and
$\bar\psi=\psi^\dagger\gamma^0$.  The momentum cutoff $\Lambda$ is
naturally related to the lattice spacing via $\Lambda\simeq \pi/2a$.
The fields $\psi$, $\psi^\dagger$ are four-component spinors,
\begin{equation}
\psi_\alpha=\left(\begin{array}{c} c_{e1\alpha} \\c_{o1\alpha}\\
c_{o2\alpha}\\c_{e2\alpha}\end{array}\right),
\qquad \psi^\dagger_\alpha=\left(c^\dagger_{e1\alpha}\,\, 
c^\dagger_{o1\alpha}\,\, c^\dagger_{o2\alpha} \,\,
c^\dagger_{e2\alpha}\right),
\label{psidef}
\end{equation}
where $1,2$ labels the Fermi point and $e,o$ labels fields on the
even and odd lattices, respectively.  The $4\times4$ $\gamma$ matrices
satisfy a Clifford algebra corresponding to the Minkowskian metric
$g^{\mu\nu}=\mbox{diag}(1,-1,-1)$. The following representation for the 
$\gamma$ matrices has been chosen:
\begin{eqnarray}
\gamma^0&=&
\left(\begin{array}{cc} \sigma_3 & 0 \\ 0 & -\sigma_3\end{array}\right),
\quad
\gamma^1=\left(
\begin{array}{cc} i\sigma_1 & 0 \\ 0 & -i\sigma_1\end{array}
\right),\nonumber\\
\gamma^2&=&\left(
\begin{array}{cc} i\sigma_2 & 0 \\ 0 & -i\sigma_2\end{array}
\right),\nonumber
\end{eqnarray}
where $\sigma_i$ are Pauli matrices, acting on the even 
and odd site fermion operators.

Expanding the AF Heisenberg interaction (\ref{Hheis}) around the two
Fermi points, we obtain
\begin{eqnarray} 
H_I&=&4J_ea^2\int\limits_{p_1,p_2,k_1,k_2\leq \Lambda}
(2\pi)^2\delta(\vec p_1+\vec k_1-\vec k_2-\vec p_2)\nonumber\\
&\times&
[c^\dagger_{e1\alpha}(\vec p_1,t)
c_{e1\beta}(\vec p_2,t)
+c^\dagger_{e2\alpha}(\vec p_1,t)c_{e2\beta}(\vec p_2,t)]\nonumber\\
&\times&
[c^\dagger_{o1\gamma}(\vec k_1,t)c_{o1\delta}(\vec k_2,t)
+c^\dagger_{o2\gamma}(\vec k_1,t)c_{o2\delta}(\vec k_2,t)]\nonumber\\
&\times&(\delta_{\alpha\delta}\delta_{\gamma\beta}-
\delta_{\alpha\beta}\delta_{\gamma\delta}/2),
\label{Hinteq1}
\end{eqnarray}
where the SU(2) Fierz identity has been used,
$\vec \sigma_{\alpha\beta}\cdot \vec \sigma_{\gamma\delta}=
2\delta_{\alpha\delta}\delta_{\gamma\beta}
-\delta_{\alpha\beta}\delta_{\gamma\beta}$.
Using
Eq.~(\ref{psidef}), Eq.~(\ref{Hinteq1}) can be written as
\begin{eqnarray}
H_I&=&-J_ea^2\int\limits_{p_1,p_2,k_1,k_2\leq \Lambda}
(2\pi)^2\delta(\vec p_1+\vec k_1-\vec k_2-\vec p_2)
\nonumber\\
&\times&
(\bar\psi_\alpha \psi_\beta \bar\psi_\gamma \psi_\delta
-\psi^\dagger_\alpha \psi_\beta \psi^\dagger_\gamma \psi_\delta)
(\delta_{\alpha\delta}\delta_{\gamma\beta}-
\delta_{\alpha\beta}\delta_{\gamma\delta}/2).\nonumber\\
\label{Hinteq2}
\end{eqnarray}
Subsequently, it is straightforward to show that Eq.~(\ref{Hinteq2})
together with Eq.~(\ref{Lkdef}) gives rise to the action $S=S_k+S_I$,
with
\begin{eqnarray}
&&S_I=J_ea^2\int dt\int\limits_{p_1,p_2,k_1,k_2\leq \Lambda}
(2\pi)^2\delta(\vec p_1+\vec k_1-\vec k_2-\vec p_2)\nonumber\\
&&\times\biggr\{
\sum_{A=1}^3\left[(\bar\psi\tau^A\psi)^2
-(\psi^\dagger\tau^A\psi)^2\right]
-\frac{(\bar\psi\psi)^2}{4}+\frac{(\psi^\dagger\psi)^2}{4}
\biggr\},\nonumber\\
\label{gnfteq1}
\end{eqnarray}
with $\tau^A$ the generators of the SU(2) symmetry with ${\rm
  Tr}\,(\tau^A\tau^B)=\delta^{AB}$ ($\tau^A=\sigma^A/\sqrt{2}$), and 
$\bar\psi\tau^A\psi=\sum_{\alpha,\beta}\bar\psi_\alpha
\tau^A_{\alpha\beta}\psi_\beta$.  The
action (\ref{gnfteq1}) is invariant under global SU(2)$\times$U(1)
corresponding to the spin orientation symmetry and total spin
conservation. Moreover the action is invariant under the discrete
transformations; space reflection, parity, and the combined CT 
(charge-conjugation and time-reversal) transformations. Naturally, it is
invariant under continuous rotations in space. However, the terms of
the form $\psi^\dagger\psi$ are not invariant under Lorentz boosts and
therefore the action is not relativistic invariant.

In what follows, we shall show that the action (\ref{gnfteq1}) lies in
the same universality class as that of a Lorentz-covariant SU(2)
invariant (2+1)-dimensional 4F model for two massless fermion flavors:
\begin{equation}
S=\int d^3x\,\left[\bar\psi i\hat \partial\psi+\frac{c G }{2}
\sum_{A=1}^3(\bar\psi\tau^A\psi)^2\right],\label{afsu2FF}
\end{equation}
where $x_0=ct$, $\hat \partial=\gamma^\mu\partial/\partial x^\mu$, 
and where $G$ is an attractive four-fermion coupling,
$cG/2\simeq J_ea^2$.  The universality only holds close to a critical
point or ultraviolet stable fixed point $G_c$ of a second-order
phase transition of Eq.~(\ref{afsu2FF}) at which the SU(2) symmetry is
dynamically broken to a U(1) symmetry.  There is a dynamical
generation of a parity-conserving ``mass term'' $m_s$ connected with
the appearance of a nonzero vacuum condensate $\langle \bar\psi \tau^3
\psi\rangle$.  This Lorentz-invariant condensate describes a staggered
spin expectation value giving rise to AF ordering
\cite{ma90a,kile99}.  In the broken phase, two massless pions or spin
waves appear as Goldstone bosons \cite{ma90a}.

The spinon propagator $S_{\alpha\beta}$ can be written as
\begin{eqnarray}
S_{\alpha\beta}(p)=\frac{\hat p A(p)\delta_{\alpha\beta}
+\sqrt{2} m_s\tau^3_{\alpha\beta}}{A^2(p)p^2-m^2_s}, \nonumber
\end{eqnarray}
with Minskowskian momentum $p^2=p_0^2-\vec p\cdot \vec p$ and where
$A(p)$ is the fermion wave function.  In the Hartree-Fock
approximation, the equation for $m_s$ gets contribution only from the
tad pole diagram and the fermion wave function $A(p)=1$. Setting
$c=1$, the gap equation for $m_s$ reads
\begin{eqnarray}
m_s=G i \int_M\frac{d^3p}{(2\pi)^3}\,\frac{4m_s}{p^2-m_s^2},
\nonumber 
\end{eqnarray}
with the subscript $M$ denoting the Minskowskian metric with cutoff
$\Lambda$.  This gap equation gives rise to the familiar critical
coupling $g^{(s)}_c\equiv 2G_c \Lambda/\pi^2=1$.  Above the critical
coupling $g=2G\Lambda/\pi^2>g^{(s)}_c$ the SU(2) spin symmetry is
broken and a $\langle\bar\psi\tau^3\psi \rangle$ condensate is formed.

Now let us show that the nonrelativistic interaction terms in
Eq.~(\ref{gnfteq1}) are irrelevant close to $g^{(s)}_c$.  We
investigate the generation of a mass $m_u$ connected with the uniform
spin expectation value $\langle \psi^\dagger \tau^3 \psi\rangle$
\cite{kile99}.  In the Hartree-Fock approximation, keeping only the
Lorentz-noninvariant terms, the spinon propagator is of the form
\begin{eqnarray}
S_{\alpha\beta}(p)&=&(\hat p\delta_{\alpha\beta}-
m_u\gamma_0\tau^3_{\alpha\beta})K_{\alpha},\nonumber\\
K_{\alpha}&=&\left\{[p_0-(-1)^{\alpha+1}
m_u/\sqrt{2}]^2-p_1^2-p_2^2\right\}^{-1},\nonumber
\end{eqnarray}
with $\alpha=1,2$ ($\alpha=\uparrow,\downarrow$).
The gap equation for $m_u$ then reads
\begin{eqnarray}
\frac{m_u}{4 G i}=
\int_M\frac{d^3p}{(2\pi)^3}\,\biggr[\frac{m_u}{2}(K_{1}+K_{2})
-\frac{p_0}{\sqrt{2}}(K_{1}-K_{2})\biggr].\nonumber
\end{eqnarray}
The bifurcation equation is
\begin{eqnarray}
\frac{1}{4Gi}=-\int_M\frac{d^3p}{(2\pi)^3}\,
\frac{p_0^2+p_1^2+p_2^2}{(p_0^2-p_1^2-p_2^2)^2},\nonumber
\end{eqnarray}
giving rise to a critical coupling $g^{(u)}_c\equiv 2G_c
\Lambda/\pi^2=3$.  This shows that when $g$ is close to the
critical value $g_c^{(s)}$, it is far ($|g-g_c^{(u)}|/g\simeq 2$) 
from the critical regime of $g^{(u)}$.  Upon increasing $g$ it first encounters the
critical value for staggered magnetization.  Hence, the
nonrelativistic terms in Eq.~(\ref{gnfteq1}) are irrelevant in that
area of the coupling constant space.

Naturally, the Hartree-Fock approximation ignores 4F fluctuations and
therefore gives a rather crude description of the critical behavior.
Nevertheless the irrelevance of the non-Lorentz-covariant terms in
Eq.~(\ref{gnfteq1}) to $g_c^{(s)}$ can be demonstrated in more
advanced approximations, such as the $1/N$ expansion.  In particular,
the SU(2) Heisenberg antiferromagnets can be generalized to SU($N$)
and studied in the $1/N$ expansion \cite{afma88maaf89}.  This is
analogous to the SU($N$) generalization of Eq.~(\ref{afsu2FF}).
Contrary to models of the Gross-Neveu type
\cite{kiya90rowapa91hakoko93}, such an expansion resembles the
topological $1/N$ expansion of 't Hooft \cite{tho74}, corresponding to
an expansion in planar Feynman diagrams instead of Fermion loops.

\begin{figure}[ht]
\resizebox*{1\columnwidth}{!}{\includegraphics{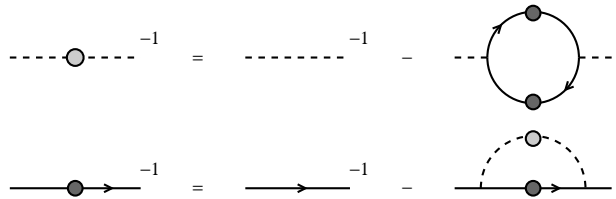}}
\caption{The leading large-$N$ truncation or ladder approximation
for the Schwinger-Dyson equations for
the spin-wave propagator $\Delta^A_\sigma(p)$ (dashed line with blob) 
and the spinon propagator $S(p)$ (solid line with blob).}
\label{fig_trunc_scheme}
\end{figure}

Recently, the generalization of Eq.~(\ref{afsu2FF}) to $N$ fermion
flavors with a SU($N$)$\times$U(1) invariant 4F potential containing
$N^2-1$ terms has been studied in the planar large-$N$ approximation
\cite{re01a}.
After a Hubbard-Stratonovich transformation
($\sigma^A=-G\bar\psi\tau^A\psi$), the action (\ref{afsu2FF}) can be
expressed as
\begin{eqnarray}
S=
\int d^3x\,\biggr\{\bar\psi i\hat \partial\psi
-\sum_A\bar\psi\tau^A\psi\sigma_A
-\frac{1}{2G}\sum_A\sigma_A^2
\biggr\}.\label{hubstrat_action}
\end{eqnarray}
In the SU($N$) case, the spin label $A$ runs from $1$ to $N^2-1$
and the fermion label from $1$ to $N$.  Thus there are $N^2-1$
composite operators, giving rise to $N^2-1$ spin propagators (connected),
$i\Delta^A_\sigma(q)\equiv\langle\sigma_A(q)\sigma_A(-q)\rangle_c$, and
$N$ spinon propagators, $iS(p)\equiv \langle \psi(p)\bar\psi(-p)\rangle$.
This allows for a 't Hooft topological $1/N$ expansion as is
conjectured in Ref.~\cite{re01a}.
Moreover, it was argued in Ref.~\cite{re01a} that the 
leading large-$N$ or planar approximation reduces to 
the ladder approximation for the so-called Yukawa vertex 
({\em i.e.,} 
$\Gamma^A_\sigma(k,p)=\tau^A{\bf 1}$).
The Yukawa vertex $\Gamma^A_\sigma$ is the fully amputated 
three-point interaction vertex for the action (\ref{hubstrat_action}).
In the ladder approximation, the Schwinger-Dyson equations for
the spin propagators $\Delta^A_\sigma$ and the spinon 
propagators $S$ form a closed set, as depicted in 
Fig.~\ref{fig_trunc_scheme}.

For the physically relevant case ($N=2$), 
the propagators $\Delta^1_\sigma$, $\Delta^2_\sigma$ 
of the auxiliary fields $\sigma_1$, $\sigma_2$
describe the Goldstone modes, whereas the field $\sigma_3$ acquires a
nonzero, but Lorentz-covariant, vacuum
expectation value in the broken phase 
($\langle\sigma_3\rangle\not =0$, $g> g_c^{(s)}$), describing the
N\'eel state, corresponding to the symmetry-breaking pattern
SU(2)$\rightarrow$U(1).  
In the subcritical region ($g\leq g_c^{(s)}$) the
propagators of all three auxiliary fields describe Goldstone precursor
modes that come down in energy as the transition is approached.
Hence, the staggered spin fluctuations or spin waves
are described by the propagators of the auxiliary fields $\sigma_A$,
which become light close to the critical point $g=g_c^{(s)}$. 

In Ref.~\cite{re01a}, the Schwinger-Dyson equations represented in 
Fig.~\ref{fig_trunc_scheme} were solved.
It was shown that the fermion or spinon propagator $S$
acquires an anomalous dimension $\zeta$ via the fermion wave function
$A(p)\sim (\Lambda/p)^\zeta$, so that at the critical coupling $G=G_c$
the fermion (spinon) propagator scales as
\begin{equation} 
S^{-1}(p)\simeq \hat p (-\Lambda^2/p^2)^{\zeta/2}.\label{Sfullasym}
\end{equation}
The dependence of the anomalous dimension $\zeta$ on $N$ is
determined, and for $N=2$, we have $\zeta\approx 0.21$ \cite{re01a}.
Moreover the model described in Ref.~\cite{re01a}, with the
dimensionless 4F coupling $g=2G\Lambda/\pi^2$, has an
ultraviolet stable fixed point at $g=g_c=1+2\zeta$. The appearance of
a positive anomalous dimension for $S$ is considered to provide a
description of the pseudogap phenomenon
\cite{frte00,khpa01,guloqush00}.

An important point is whether the critical coupling $g_c^{(s)}\simeq
1+2\zeta$ is in agreement with the estimations for the physical
parameters.  Since $G c/2\simeq J_ea^2$, $c\simeq 2|\chi| a$,
$\Lambda\simeq \pi/2a$, we have that $g\simeq J_e/\pi |\chi|$.  With
the estimation $g^{(s)}_c\simeq 1.4$, we obtain that a value
$|\chi|\approx 0.23 J_e$ would get us close to criticality.  This is
remarkably close to the mean-field value $|\chi|\approx 0.24 J $ given
in Ref.~\cite{afma88maaf89}.  

In the ladder approximation (Fig.~\ref{fig_trunc_scheme}), 
the connected propagator of the
$\sigma_A$ field in momentum space reads
\begin{eqnarray}
\Delta_\sigma^{A-1}(q)\equiv
-\frac{1}{G}+i\int_M d^3k\,
{\rm Tr}\,\left[\tau^A S(k+q)\tau^A S(k)\right],\nonumber
\end{eqnarray}
where $S(k)$ is the full spinon propagator, given by 
Eq.~(\ref{Sfullasym}),
see also Fig.~\ref{fig_trunc_scheme}.
In the subcritical or SU($N$) symmetric regime, we can take $S(k)=\hat
k/[k^2 A(k)]$.  The integral can be performed, and
$\Delta_\sigma^A(q)$ has the following scaling form for $|q|\ll
\Lambda$ ($q^2=q_0^2-q_1^2-q_2^2$):
\begin{eqnarray}
\Delta_\sigma^{A}(q)&\simeq& -\frac{C}{\Lambda}
\frac{(-\Lambda^2/q^2)^{\zeta+1/2}}{
\left[1+\left(-m_\sigma^2/q^2\right)^{\zeta+1/2}\right]},
\label{delseq}
\end{eqnarray}
where $C$ is some flavor dependent positive constant \cite{re01a}.
The mass $m_\sigma$ (spin-wave stiffness) denotes the position of the
resonance peak given by the imaginary part of Eq.~(\ref{delseq}), and
plays the role of the inverse correlation length \cite{guhare01},
\begin{equation}
m_\sigma/\Lambda\sim \left( g_c^{(s)}-g \right)^{1/(1+2\zeta)}.
\label{resonance}
\end{equation}
From these expressions, it follows that the critical exponents $\eta$,
$\nu$, and $\gamma$ are $\eta=1-2\zeta$, $\nu=1/(1+2\zeta)$, and
$\gamma=1$.  These exponents satisfy the three-dimensional
hyperscaling equations \cite{re01a}.  Moreover, in Lorentz-invariant
field theories, the scaling of the energy equals the scaling of
momentum, and consequently the dynamical scaling exponent $z=1$.

\begin{figure}[h]
\rotatebox{-90}{
\resizebox*{0.6\columnwidth}{!}{\includegraphics{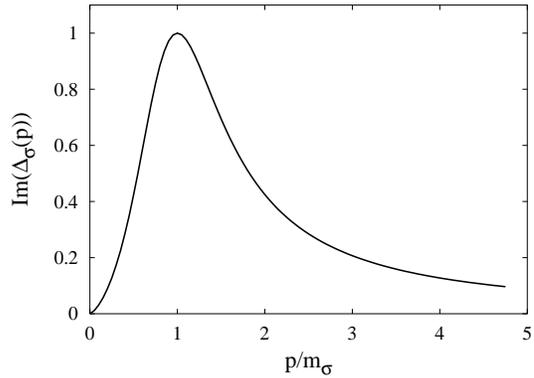}}}
\caption{The spin susceptibility or spin-correlation function 
${\rm Im}\,[\Delta^A_\sigma(p)]$ 
vs $p/m_\sigma$, with the peak renormalized at unity, $\zeta=0.21$ 
($\eta=0.58$).}
\label{fig_res}
\end{figure}

In experiments, the isotropic dynamical susceptibility
$\chi^{\prime\prime}(q)$ is measured \cite{fongetal}, with $q=(\omega,
\vec q)$.  In Ref.~\cite{fongetal} magnetic resonances were observed
in underdoped and optimally doped YBa$_2$Cu$_3$O$_{6+x}$, with the
famous 41-meV peak at optimal doping.  For lower doping the resonance
peak shifts to lower energies.  It is tempting to assume that, along
the lines of Refs.~\cite{kile99,guhare01}, these resonances might be
described by the Dirac models, see Fig.~\ref{fig_res}.  
However, optimal doping ($\delta\simeq
0.2$) is rather far from N\'eel doping, which is one order of
magnitude less, $\delta_c \simeq 0.02$ \cite{brle01}.  Therefore the
question is whether we are ``sufficiently'' close to the scaling
region of the N\'eel transition.  

Let us compare a couple of (2+1)-dimensional Lorentz-covariant quantum field
models, capable of describing AF fluctuation and N\'eel
transition.  For instance, Kwon \cite{kwon} presented a nodal d-wave
spin liquid model for the N\'eel transition, which has the universality
class of the Gross-Neveu model.  This model has a single AF order
parameter, corresponding to the singlet composite order parameter
$\langle\bar\psi\psi\rangle$, with a single two-component Dirac
fermion ($N=1$).  The universality class of the model presented in
this paper ({\em i.e.,} Eq.~(\ref{afsu2FF})) deviates considerably
from the Gross-Neveu model \cite{re01a}.  Although the anomalous
dimension $\eta$ for both models is comparable: $\eta\approx 0.58$ for
the present model, and $\eta=16/(3\pi^2 N)\approx 0.54$ ($N=1$) for
the Gross-Neveu model, the main difference is in the value of the
anomalous dimension of the fermion propagator, which for the present
model is $\zeta\approx 0.21$, whereas for the Gross-Neveu model it is
$\eta=2/(3\pi^2N)\approx 0.07$ ($N=1$).  Another approach was adopted
in Ref.~\cite{chsaye94}, where the antiferromagnetic correlations and
the N\'eel transition are described by the nonlinear sigma model in
the large-$N$ expansion.  In that work, the order parameter is a
three-component one, corresponding to the three spin-1 components, and
no Dirac fermions are taken into account.  The anomalous dimension
$\eta$ for the spin-waves turned out to be $\eta=8/(3\pi^2 N)\approx
0.09$ ($N=3$), which is considerably smaller than $\eta$ for the two
above mentioned 4F Dirac models.  In this paper, we have a 
three-component order parameter (the composite spin degrees) and 
two flavors ($N=2$) of four-component Dirac fermions.  To determine which
universality class describes the N\'eel transition, the low doping
region $\delta\sim\delta_c$ needs to be examined in more detail
experimentally.

Since in the subcritical phase, all three correlation functions of the
staggered spin components are degenerate, the imaginary part of the
correlation function $\Delta_\sigma^A(q)$ is directly proportional to
the so-called odd acoustic mode of $\chi^{\prime\prime}(q)$
\cite{brle01}.  The spin-correlation function only gets low-energy
contributions from the staggered spin operators;
\begin{eqnarray}
\langle S_A(p) S_B(-p)\rangle&\propto&
\langle \sigma_A(p) \sigma_B(-p)\rangle
\simeq\delta_{AB} i\Delta_\sigma^A(p).\nonumber
\end{eqnarray}
Moreover, since $g$ is proportional to $J_e$, $g$ reduces when doping is
increased \cite{lesa99an01}.  The critical coupling $g_c^{(s)}$ of the
N\'eel transition is a quantum critical point \cite{sa00} and
corresponds to a critical N\'eel doping $\delta_c\ll 1$.
Equation~(\ref{resonance}) gives the relation between the position of
the peak of the magnetic resonance and the doping rate $\delta$.
Assuming $\delta_c$ is sufficiently close to zero, we obtain that the
resonance peak is linearly proportional to doping (for small doping
rates with $\delta>\delta_c$). Consequently, the peak position moves
to lower energies when doping is reduced; near $\delta_c$ the peak
height diverges. For doping values $\delta<\delta_c$ spin waves
appear.

In summary, we have shown that a (2+1)-dimensional Lorentz-invariant 
4F model with a global spin SU(2) symmetry describes the low-energy 
time-dependent ``quasi-particle'' spin excitations of the t-J model 
near the AF wave vector. The spin excitations are given in terms of 
quantum fluctuations around the mean-field $\pi$-flux phase. The N\'eel
transition is described as the DSB of SU(2)$\rightarrow$U(1) in the
model. The magnitude of the critical coupling $g^{(s)}_c$ of the 4F
model turned out to be in good agreement with the input parameters $J$
and $|\chi|$, which define the flux-phase spin liquid. Nevertheless,
the question of the precise effects of hole doping on the AF
interaction and the anisotropy of the Dirac spectrum is left open.
For the future, it would be interesting to include the contribution of
the slave bosons (holons) ({\em e.g}, see Refs.~\cite{kile99,brle01})
in order to take into account anisotropy and to determine the
effective AF coupling $J_e$. We demonstrated that the unstable spin
modes found in experiments might be well described by the
Nambu-Goldstone boson precursor modes in the subcritical region.
Furthermore, the spinon propagator acquires an anomalous dimension,
but remains gapless near the AF wave vector in the normal state. The
appearance of a spinon anomalous wave function gives a natural
description of the pseudogap phenomenon.

\acknowledgments{The author thanks V.P.~Gusynin for useful suggestions.}
\end{document}